# On the quantum principles of cognitive learning

Alexandre de Castro
Laboratório de Matemática Computacional – CNPTIA/Embrapa, Campinas
13083-886 SP, Brazil. alexandre.castro@embrapa.br

**Abstract:** Pothos & Busemeyer's query as to whether or not quantum probability can provide a foundation for the cognitive modeling embodies so many underlying implications that the subject is far from exhausted. In this brief commentary, however, I suggest that it is possible, even likely, to find a quantum statistics to describe the cognitive behavior, especially, with regard to the conceptual schema of meaningful learning.

The principles of superposition and entanglement are central to quantum physics. Quantum superposition is commonly considered to be a mapping of two bit states into one. Mathematically, we can say that it is nothing more than a linear combination of classical (pure) states. As to the quantum entanglement, it refers to a short- or long-range operation in which a strongly correlated state, a mixed state, is built from pure states. An important feature of this mixed state is that it cannot be represented by a tensor product of states, and, once such an entangled system is constructed, it cannot be dissociated (Dirac 1999).

In this target article, Pothos & Busemeyer elegantly argue that there may be quantum principles – notably superposition and entanglement – at play in the context of human cognitive behavior. They draw attention to the pertinent idea that the concept of quantum likelihood can provide a novel guidance in cognitive modeling. Viewed in these terms, I agree with Pothos & Busemeyer, because I, too, have identified both superposition and entanglement from the cognitive premises formulated within the concept of subsumption (assimilation) of information proposed by Ausubel (1963; 1968).

From the point of view of the process of subsuming information, the material meaningfully incorporated within an individual's cognitive structure is never lost, but a state called "forgetting" takes place in a much more spontaneous manner, because it is a continuation of the very process of associative subsumption by which one learns. This forgetting state is an obliterative stage of the subsumption, characterized by Ausubel as "memorial reduction to the least common denominator" (Brown 2000). This "memorial reduction" required for the acquisition of new meanings (knowledge) is clearly (and remarkably) a conceptual model of quantum superposition of mental states, and, consequently, this cognitive behavior can be generically expressed by a quantum operation of retention of information, a cognitive squeeze, as follows: |bit>+|bit>=|qubit> . In addition, Ausubel (1963) claimed that, when the obliterative stage of subsumption begins, "specific items become progressively less dissociable as entities in their own right until they are no longer available and are said to be forgotten".

This "forgetting" theorized by Ausubel seems to reflect very well the entanglement included in the central idea of quantum cognition raised by Pothos & Busemeyer. In passing, Vitiello's work (1995) quoted in the target article also addressed the squeeze of information and the forgetting dynamics to describe the cognitive behavior, although that work does not properly refer to Ausubelian subsumption of information. Nevertheless, more in line with Ausubel's premises, Brookes' pioneer work (1980) on the cognitive aspects in the information sciences (Bawden 2008) provides a quantitative sharp bias of meaningful learning, albeit seldom examined from this perspective (Neill 1982 ; Cole 1997; Cole 2011). Most of the works found in the literature quote Brookes' fundamental equation of information science, $K(S)+I \Rightarrow K(S+\Delta S)$, – here assumed as an obliterative synthesis that exhibits short-term instability –, as merely a shorthand description of knowledge transformation, wherein the state of mind $K(S)$ changes to another state $K(S+\Delta S)$, because of an input of information $I$, being $\Delta S$ an indicator of the effect this transformation (Cornelius 2002 ; Bawden 2011).

On the other hand, in another seminal article published in the early 1980s and entitled, "Information technology and the science of information," Brookes (1981) – although in an incipient approach – conjectured outright that the recipient knowledge structure included in the fundamental equation could be quantitatively treated, which, in a subjacent manner, links his work to the assimilative schema of information expressed by the Ausubelian symbolic quantities, even though these quantities are situated in an semi-quantitative pictorial landscape (Moreira 2011; Seel 2012). In support of this idea, Todd (1999), in an important review article published at the end of the last century, also advocated that the unit of information embedded in Brookes' theory is a concept derived from Ausubel's learning theory. For such reasons, I am convinced that Brookes' equation faithfully shapes the Ausubelian retention of information, or, more specifically, the superposition and entanglement of information underlying the subsumption.

Interestingly, if we take into account that information is the boundary condition of the human cognitive system – and if we continue to perceive knowledge from a Nietzschean standpoint, in which subject and object are confused – then the reciprocal reckoning of Brookes' equation, $K(S+\Delta S) \rightarrow K(S)+I$, in addition to providing a typical scenario of information retention, also seems to give us a symbolic (and quantum) translation of Jose Ortega y Gasset's famous maxim, "I am I plus my circumstances," which Gasset (1998) placed at the metaphysical core of his epistemological approach of perspectivism.

Brookes himself addressed a peculiar notion of perspectivism in his work. In a pioneering way, Brookes (1981) suggested in a reductionist geometric context – although without clarification – a rough sketch, a skeleton, of a logarithmic equation to represent the carrying of information into the human mind on the same basis as Hartley's law (Seising 2010), seeing that Hartley's law – predecessor to Shannon's idea of channel capacity – had, up to that time, been designed solely to handle information in a purely physical system. However, albeit Brookes has made a valuable contribution by suggesting a Hartley-like behavior for information processing in the mind, he was not able to identify the appropriate cognitive variables for the implementation of his physicalistic approach from the perspectivism.

As an alternative to Brookes' approach, I showed in a recent preprint (Castro 2012) that the conceptual schema of meaningful learning leads directly to a Shannon-Hartley-like model (Gokhale 2004), and that this model can be interpreted from basilar cognitive variables, such as information and working memory. Moreover, starting this learning schema, I have found that the ratio between the two mental states given by the Brookes' fundamental equation of information science is as follows: $\dfrac{K(S)}{K(S+\Delta S)} \geq e^{-\frac{\Delta E}{k_B T}}$, where $\Delta E$ is the free energy of the ensemble, $k_B$ is Boltzmann's constant, and $T$ is the absolute temperature. The so-called Boltzmann factor, $e^{-\frac{\Delta E}{k_B T}}$, is a weighting measure that evaluates the relative probability of a determined state occurring in a multi-state system (Carter 2001), that is, it is a "non-normalized probability" that needs to be "much greater" than unity ($\gg 1$) for the ensemble to be described for non-quantum statistics; otherwise, the system exhibits quantum behavior.

As a result, this calculation shows that the internalization of one unit of information into an individual's mental structure gives rise to a Boltzmann "cognitive" factor $\leq 2$, which provides us a circumstantial evidence that the subsumption of new material, as a cognitive process, requires a quantum-statistical treatment, such as Pothos & Busemeyer have proposed.